\documentclass[12pt]{article}
\usepackage{amsmath,amssymb}
\usepackage{graphicx,psfrag,epsf}
\usepackage{enumerate}
\usepackage{natbib}
\usepackage{url} 

\newcommand{\blind}{0}

\begin{document}

\def\spacingset#1{\renewcommand{\baselinestretch}%
{#1}\small\normalsize} \spacingset{1}


\if0\blind
{
  \title{\bf Mastering the body and tail shape of a distribution}
  \author{Matthias Wagener\thanks{This work is based on the research supported in part by the National Research Foundation of South Africa (SARChI Research Chair- UID: 71199; and Grant ref. CPRR160403161466 nr. 105840). Opinions expressed and conclusions arrived at are those of the author and are not necessarily to be attributed to the NRF.}\hspace{.2cm}\\
    Department of Statistics, University of Pretoria\\\\    
    Andriette Bekker \\
    Department of Statistics, University of Pretoria\\
    and\\
    Mohammad Arashi\\
    Department of Statistics, University of Pretoria}
  \maketitle
} \fi

\if1\blind
{
  \bigskip
  \bigskip
  \bigskip
  \begin{center}
    {\LARGE\bf Mastering the body and tail shape of a distribution}
\end{center}
  \medskip
} \fi

\bigskip
\begin{abstract}
The normal distribution and its perturbation has left an immense mark on the statistical literature. Hence, several generalized forms were developed to model diﬀerent skewness, kurtosis, and body shapes. However, it is not easy to distinguish between changes in the relative body and tail shapes when using these generalizations. What we propose is a neat integration approach generalization which enables the visualization and control of the body and the tail shape separately. This provides a ﬂexible modeling opportunity with an emphasis on parameter inference and interpretation. Two related models, the two-piece body-tail generalized normal and the two-piece tail adjusted normal are swiftly introduced to demonstrate this inferential potential. The methodology is then demonstrated on heavy and light-tailed data.
\end{abstract}

\noindent%
{\it Keywords:} generalized normal, body-tail, kurtosis, inferential statistics, Bitcoin
\vfill

\newpage
\spacingset{1.45} 
\section{Origins}
\label{sec:origins}

Flexible modeling is an ongoing study in distribution theory that dates back as far as 1879 when Galton pioneered the log-normal distribution \cite{deVries1894}. Since then, the field has exploded with new distributions and ways of generating them. These models include finite mixture models \cite{mclachlan2000}, variance-mean mixtures \cite{barndorff82}, copulas \cite{nelsen07}, the Box-Cox transformation \cite{box64}, order-statistics-based distributions \cite{jones04}, probability integral transformations of \cite{ferreira06}, and the Pearson system of distributions \cite{johnson94}, to name but a few. The impact of flexible modeling is further underscored by their successful integration into classical statistical approaches such as time series analysis \cite{hansen94}, space-state models \cite{naveau05}, random fields \cite{allard07}, regression models \cite{azzalini08}, linear mixed effects models \cite{linearmixed05}, non-linear mixed-effects models \cite{nonlinearmixed19}, Bayesian statistics \cite{rubio14}, and Bayesian linear mixed models \cite{baylinmod18}. 

In \cite{ley14flex} and \cite{jones15}, the respective authors formulate some of the desirable traits of a univariate flexible model. We focus on three highlights the authors have in common:

\begin{itemize}
	
	\item A finite number of well interpretable parameters:  
	These include parameters that specifically control location, scale, skewness and kurtosis. 
	
	\item Favorable estimation properties: 
	It is important the parameters can be estimated correctly to ensure correct predictions and inferences from the model. Inferentially speaking the ideal would be to have a model to use in tests of normality.
	
	\item 
	Simple tractability: Closed form expressions are still desirable despite modern computational power. Simple formulae describing characteristics of distributions aid in exposition and additionally improve computational implementation and speed.
	
\end{itemize}

A specific generalization of the normal is relevant to this contribution. It has many names such as: the exponential power (EP), generalized power, generalized error, generalized Gaussian, and generalized normal (GN) distribution. This family is originally proposed by \cite{subbotin23} and later on again by \cite{box62} and \cite{box73}. A more complete review of this is given by \cite{nadarajah12}. The GN has been generalized to accommodate skew data in many different ways. These skewed GN distributions are summarized in Table \ref{tab:sgn}.

\begin{table}[h!]
	\caption{The GN and its skew generalizations.}
	\centering
	\resizebox{0.6 \textwidth}{!}{	
		\begin{tabular}{@{\extracolsep{5pt}} ll} 
			\\[-1.8ex]\hline 
			\hline \\[-1.8ex] 
			Author & Name of distribution\\ 
			\hline
			1986 Azzalini \cite{azzalini08}  &Skew EP \\
			1995 Fernandez, Osiewalski, and Steel \cite{fernandez95} & Skewed EP  \\
			2003 Ayebo and Kozubowski \cite{ayebo03} & Asymmetric EP  \\
			2007 Komunjer \cite{komunjer07}& Asymmetric Power Family \\
			2009 Zhu and Zinde-Walsh \cite{zhu09}& Generalized Error Class\\
			2011 Bottazzi and Secchi \cite{bottazzi11}& Asymmetric EP \\
			\hline \\[-1.8ex] 
		\end{tabular}
	}
	\label{tab:sgn}
\end{table}

In this paper we provide a new body-tail generalization of the normal distribution (BTGN) which also includes the GN distribution. The strong ties of the BTGN to the literature makes it an ideal foundation for achieving the aforementioned desirable traits of a flexible model.

\section{Integrating to new distributions }
\label{sec:motiv}
Given some ``appropriate'' derivative kernel function, $k'(x)$,  a new distribution can be generated by simply integrating $k'(x)$ and normalizing the resulting function to give a new density $f(x)$. 

Using integration to generate new distributions is not completely new, very recently an integration approach  is used in a reliability context, see \cite{baker19}. It however seems that this particular type of deliberate integration of a derivative kernel function has not been done before. 

\begin{figure}[ht!]
	\centering{
		\includegraphics[width=3in,keepaspectratio]{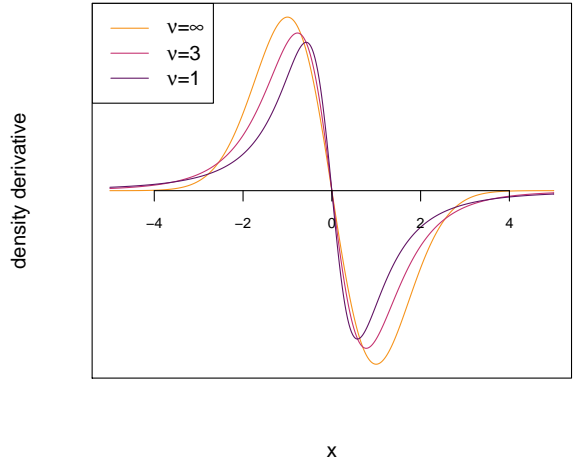}
	}
	
	\caption{The derivative of the t distribution density. Note that for heavier tails we have a lower more abrupt cross-over at the mode, faster derivative values in the body, and slower derivative values in the tails.}
	\label{fig:tderivative}
\end{figure}

The link between the derivative kernel function and the density can be exploited to define the shape of new distribution as needed. By studying the derivative of the t-distribution we know how a derivative kernel function should behave to obtain lighter and heavy tails, see Figure \ref{fig:tderivative}.

Now, the derivative of the GN distribution kernel is given below:
\begin{equation}
\frac{d}{dx}\text{e}^{-|x|^\beta}=-\beta \text{ sign}(x) |x|^{\beta-1}\text{e}^{-|x|^\beta},
\end{equation}
where $\beta>0$.

Simply replacing $|x|^{\beta-1}$ with $|x|^{\alpha-1}$, gives a more flexible derivative kernel where $\alpha$ determines body shape and $\beta$ controls the tail behavior:
\begin{equation}
\label{eq:btgnkern}
k'(x)=-\beta\text{ sign}(x) |x|^{\alpha-1}\text{e}^{-|x|^\beta},
\end{equation}
where $x\in \Re$ and $\alpha,\beta>0$.

\begin{figure}[ht!]
	\centering{
		\includegraphics[width=3in,keepaspectratio]{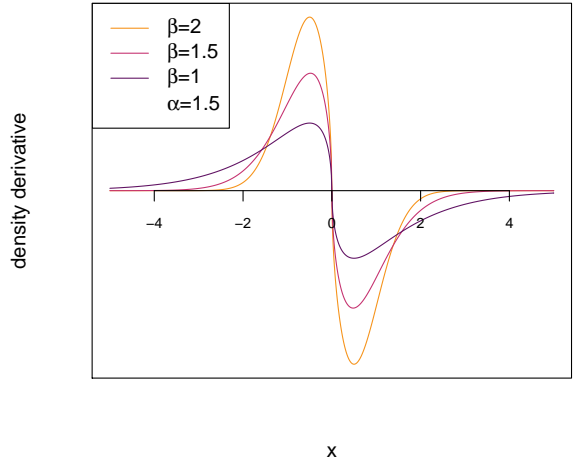}
	}
	
	\caption{The proposed derivative kernel for the BTGN distribution, see Equation (\ref{eq:btgnkern}).}
	\label{fig:gnderivative}
\end{figure}

The role of $\beta$ is very similar to $\nu$ in the t distribution but with a wider range including lighter tails, see Figure \ref{fig:gnderivative}. Interestingly, defining a derivative kernel in such a way would be proportional to the kernel of a generalized gamma distribution \cite{stacy62}.

This then, lays the foundation for the body-tail generalized normal (BTGN) distribution that will, by definition, contain the GN distribution, for $\alpha=\beta$, and have simultaneous control of the body and tail shape.
\section{The body-tail generalized normal distribution}
\label{sec:btgn}
Let the BTGN derivative kernel, $k'(x)$, be defined by Equation (\ref{eq:btgnkern}). Thus the indefinite integral of $k'(x)$, Equation (\ref{eq:indefint}), yields a new symmetric kernel Equation (\ref{eq:symkernbtgn}). 
\begin{align}
\text{For $x>0$:}\notag\\
\int k'(x)dx
&=\int- y^{\frac{\alpha-1}{\beta}}y^{\frac{1}{\beta}-1}\text{exp}\left(-y \right)dy\notag\\
&=\Gamma\left(\frac{\alpha}{\beta},x^\beta \right).
\label{eq:indefint}
\end{align}
\begin{equation}
\label{eq:symkernbtgn}
\therefore k(x)		=\Gamma\left(\frac{\alpha}{\beta},|x|^\beta \right),
\end{equation}

where $\Gamma\left(\cdot,\cdot \right)$ is the upper incomplete Gamma function \cite{Gr<3dstheyn}. 
For a bona fide density, the normalizing constant for $k(x)$ is given by solving the integral below with Lemmas 1 and 2.
\begin{align}
2\int_{0}^{\infty}\Gamma\left(\frac{\alpha}{\beta},x^\beta \right)dx
&=2\Gamma\left( \frac{\alpha+1}{\beta}\right).
\end{align}

The body-tail generalized normal distribution density is then given by:
\begin{equation}
\label{eq:btpdf}
f(x;\alpha,\beta)=\frac{\Gamma\left(\frac{\alpha}{\beta},|x|^\beta \right)}{2\Gamma\left( \frac{\alpha+1}{\beta}\right)},
\end{equation}

where $x\in\Re$ and $\alpha, \beta>0$.
Again, note that if $\alpha=\beta$ we have the regular generalized normal distribution as discussed in Section \ref{sec:origins}:

\begin{align*}
f(x;\alpha,\alpha)&=\frac{\int_{|x|}^{\infty}|x|^{\alpha-1}\text{e}^{-|x|^\alpha}dx}{\Gamma\left(\frac{\alpha+1}{\alpha}\right)}
\notag\\
&=\frac{\alpha\text{e}^{-|x|^\alpha}}{\Gamma\left(1/\alpha\right)}.
\end{align*}
This of course implies that for $\alpha=\beta=2$ we have a normal distribution with scale, $\sigma=\frac{1}{\sqrt{2}}$, and the Laplace distribution for, $\alpha=\beta=1$, respectively.

\begin{figure}[ht!]
	\centering{
		\includegraphics[width=7cm,keepaspectratio]{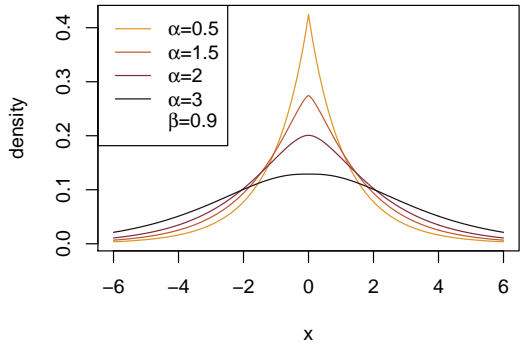}
	}
	
	\caption{The density of the BTGN distribution for different body shapes $\alpha$.}
	\label{fig:btpdf1}
\end{figure}

\begin{figure}[ht!]
	\centering{
		\includegraphics[width=7cm,keepaspectratio]{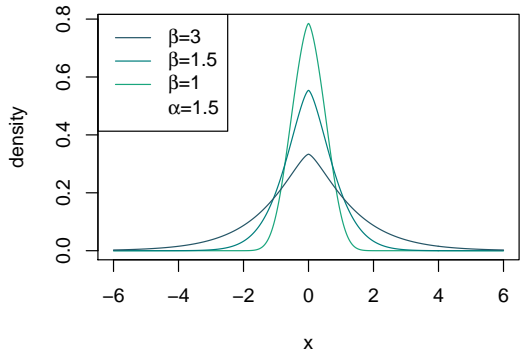}
	}
	
	\caption{The density of the BTGN for different tail shapes of $\beta$. Note that for $\beta<\alpha$ tails are heavier relative to what is usually associated with the body of the GN distribution and vice versa.}
	\label{fig:btpdf2}
\end{figure}

In Figures \ref{fig:btpdf1} and \ref{fig:btpdf2} the inferential potential of the BTGN distribution becomes clear. The simultaneous control of the body and tail shape make it possible to test the cause of deviation from normality due to body and tail shape. For instance, two other candidates that include the normal distribution come to mind, the t and generalized hyperbolic distributions. Although these models have a rich statistical literature, the former can not model lighter than normal tails and the latter has a shape and tail shape parameter that interact \cite{genhypscott11}. This gives a key advantage for using the BTGN in inference.

Next, the cumulative distribution function, $F(x)$, is easily calculated from Equation (\ref{eq:btpdf}) using Lemma 2 in the Appendix. For $x\le0$:

\begin{align*}
F(x)
&=\int_{-\infty}^{x}
\frac{\Gamma\left({\alpha}/{\beta},|t|^\beta \right)}
{2\Gamma\left(\frac{\alpha+1}{\beta}\right)}dt\notag\\
&=
2^{-1}\Gamma\left(\frac{\alpha+1}{\beta}\right)^{-1}\cdot
\int_{-x}^{\infty}\Gamma\left({\alpha}/{\beta},t^\beta \right)dt\notag\\
&=
\frac
{\Gamma\left( \frac{\alpha+1}{\beta}, (-x)^\beta\right)
	-x\cdot\Gamma\left(\frac{\alpha}{\beta},(-x)^\beta \right)}
{2\Gamma\left(\frac{\alpha+1}{\beta}\right)}.
\end{align*}
For $x>0$ the relation $F(x)=1-F(-x)$ can simply be used.

Finally, the absolute $r$'th moments are derived since the odd moments of the BTGN are zero. From Equation (\ref{eq:btpdf}), and Lemmas 1 and 2 we have that:

\begin{align*}
\label{eq:mom}
E(|X|^r)&=
2\int_0^\infty
x^r\frac
{\Gamma\left({\alpha}/{\beta},|x|^\beta\right)}
{2\Gamma\left(\frac{\alpha+1}{\beta}\right)}dx \notag\\
&= \frac
{\Gamma\left(\frac{\alpha+r+1}{\beta}\right)} 
{2(r+1)\Gamma\left(\frac{\alpha+1}{\beta}\right)}.
\end{align*}

In the following section the BTGN used in flexible modeling and inference. The location-scale BTGN PDF is thus needed and given below: 

\begin{equation}
\label{eq:btpdflocscal}
f(x;\alpha,\beta)=\frac{\Gamma\left(\frac{\alpha}{\beta},\left|\frac{x-\mu}{\sigma} \right| ^\beta \right)}{2\sigma\Gamma\left( \frac{\alpha+1}{\beta}\right)}.
\end{equation}

\section{Methodological demonstration}
The previous sections focused on the desirable traits of the BTGN such as interpretable parameters, tractability, and inferential potential. In this section we show the practical application of the BTGN as a building block for flexible models. In order to make the BTGN more applicable to typical flexible modeling situations, skewness parameter $\psi$ is added. This is done by two-piece scaling, which has an ``easy and clean set-up'' and strong parameter orthogonality \cite{jones11}. The successful Azzalini-type skewing can also be considered in later studies for their good stochastic properties, elegant generating mechanisms, and fitting properties, see \cite{deHel1909} and \cite{azzalini85}.

The scaled and shifted TPBGN is given by Equation (\ref{eq:tpbgnpdf}). The scaled and shifted two-piece tail adjusted normal (TPTAN) is sub-model in the special case of $\alpha=2$.

\begin{equation}
\label{eq:tpbgnpdf}
f(x;\mu,\sigma,\alpha,\beta,\psi)=	\left\{\begin{array}{ll}
\frac{2}{\psi+\frac{1}{\psi}}
\frac{\Gamma\left(\frac{\alpha}{\beta},\left( \frac{-x-\mu}{\sigma}\right) ^\beta \right)}
{2\sigma\Gamma\left( \frac{\alpha+1}{\beta}\right)}
& 	x\leq \mu \\
\frac{2}{\psi+\frac{1}{\psi}} 
\frac{\Gamma\left(\frac{\alpha}{\beta},\left( \frac{x-\mu}{\sigma}\right) ^\beta \right)}
{2\sigma\Gamma\left( \frac{\alpha+1}{\beta}\right)}
& 	\mu<x \\
\end{array} 
\right.
\end{equation}
Given these two flexible models, we analyze a heavy and light tailed data set inferring some of the characteristics of the data.

\subsection{Bitcoin Returns}
\begin{figure}[ht!]
	\centering{
		\includegraphics[width=3in,keepaspectratio]{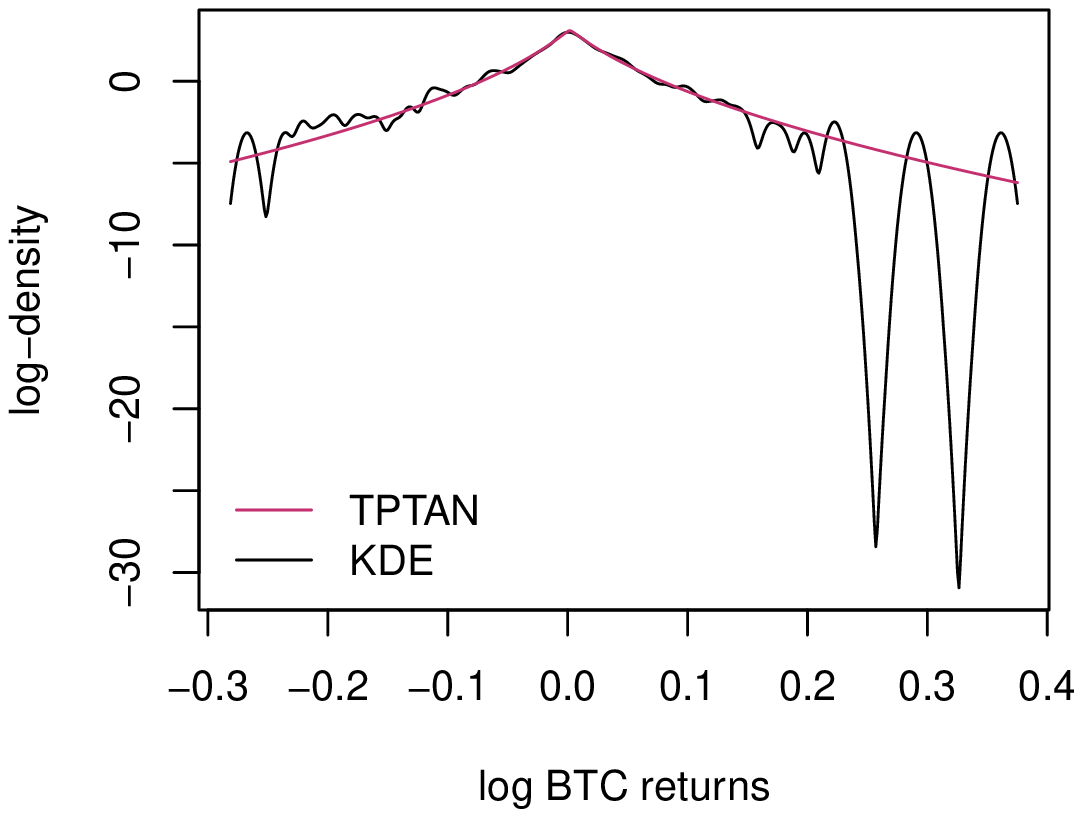}
	}
	\caption{}
	\label{fig:btcapp}
\end{figure}

In the first case the data represents the log daily returns of Bitcoin during the time period of 2013/07/07-2018/12/17 (1989 days), available at \url{https://community-api.coinmetrics.io/v2}. In Figure \ref{fig:btcapp} the kernel density estimate (KDE) and the fitted TPTAN density is shown in log scale for clarity. The approximate Bayes' factor is used to determine evidence in favor of a model, see \cite{bayf95}. The TPTAN is fitted and compared to the BTGN to infer whether the body of the distribution is non-normal. As competing models the Azzalini skew t (ST) \cite{azzalini85} and the normal inverse Gaussian distributions (NIG) \cite{barndorff78} are also fitted. From Table \ref{tab:btcappbf} we can deduce that the body of the returns data is normal-like with heavy tails and that the TPTAN represents the data best.

\begin{table}[!htbp] \centering 
	\caption{Bayes' factor and interpretation for the Bitcoin Returns Data.} 
	\begin{tabular}{@{\extracolsep{5pt}} ccc} 
		\\[-1.8ex]\hline 
		\hline \\[-1.8ex] 
		& Bayes' Factor & $H_0$ Evidence\\ 
		\hline \\[-1.8ex] 
		ST & $66.223$ & Very strong \\ 
		NIG & $23.004$ & Very strong \\ 
		TPTAN & $H_0$ &  \\ 
		TPBTGN & $12.012$ & Very strong \\ 
		\hline \\[-1.8ex] 
	\end{tabular} 
	\label{tab:btcappbf}
\end{table}

\begin{figure}[ht!]
	\centering{
		\includegraphics[width=3in,keepaspectratio]{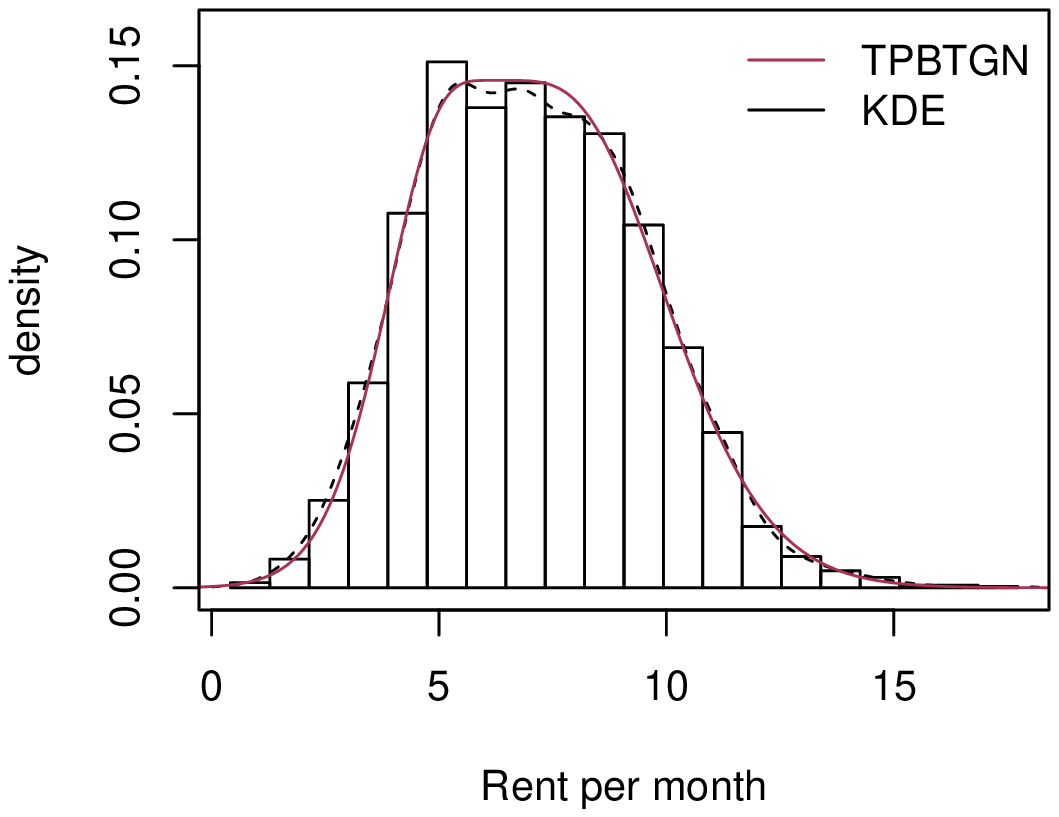}
	}
	
	\caption{}
	\label{fig:munichapp}
\end{figure}
\subsection{Munich Rent}
In the second case the data represents 3082 observations of net rental price per square meter. The data comes for a rent survey in Munich in the year 1999 \cite{rent99}, available in the R package gamlss.data. In Figure \ref{fig:munichapp} the light tailed data is presented with the fitted TPTAN density. Once again, the approximate Bayes' factor is used and both the TPBTGN and TPTAN are fitted. As competing models two group finite mixtures of normal (FMN) and Gamma (FMG) distributions are fitted. From Table \ref{tab:munichapp} we can deduce that the body of the rent data is non-normal and that the TPBTGN represents the data best.

\begin{table}[!htbp] \centering 
	\caption{Bayes factor and interpretation for the Munich Rent Data.} 
	\begin{tabular}{@{\extracolsep{5pt}} ccc} 
		\\[-1.8ex]\hline 
		\hline \\[-1.8ex] 
		& Bayes Factor & $H_0$ Evidence \\ 
		\hline \\[-1.8ex] 
		FMG & $13.615$ & Very strong\\ 
		FMN & $8.121$ & Strong\\ 
		TPTAN & $9.794$ & Strong\\ 
		TPBTGN & $H_0$ & \\ 
		\hline \\[-1.8ex] 
		\label{tab:munichapp}
	\end{tabular} 
\end{table} 
\section{Appendix}
The appendix has two lemmas for the calculations of the density, cumulative distribution function, and moments.

\subsection*{Lemma 1}
Let $\alpha,\beta>0$, then the following limit holds true below:

\begin{equation}
\label{eq:lem1}
\lim\limits_{x\to \infty} x^{k}\Gamma\left( \frac{\alpha}{\beta},x^\beta\right) =0 \text{ for } k\in\Re.
\end{equation}

\subsubsection*{Proof:}
If $k\le0$ both factors in the left-hand side of Equation (\ref{eq:lem1}) tend to zero of is finite as $x$ tends to infinity.
If $0<k$, by L'Hospital rule:
\begin{align}
&\lim\limits_{x\to \infty} x^k\Gamma\left( \frac{\alpha}{\beta},x^\beta \right)
\notag\\
&=
\lim\limits_{x\to \infty} 
\frac{x^{\alpha+\beta+k-1}}{\text{e}^{x^\beta}}\cdot
\frac{\beta}{k}
\notag\\
&=0.
\end{align}  

\subsection*{Lemma 2}
Let $\alpha,\beta>0$, then the following integral identity holds true below:

\begin{equation}
\label{eq:lem2}
\int_x^\infty x^r\Gamma\left( \frac{\alpha}{\beta},x^\beta\right)dx 
= \frac
{\Gamma\left(\frac{\alpha+r+1}{\beta},x^\beta \right)
	-x^{\frac{r+1}{\beta}}\Gamma\left(\frac{\alpha}{\beta},x^\beta \right)} 
{r+1}.
\end{equation}

\subsubsection{Proof:}
Integrating by parts, set
$v'(y)=\frac{r+1}{\beta}y^{\frac{r+1}{\beta}-1}$ and $u(y)=\Gamma\left({\alpha}/{\beta},y \right)$. This implies that $v(y)=y^{\frac{r+1}{\beta}}$ and $u'(y)=-y^{\frac{\alpha}{\beta}-1}\text{e}^{-y}$. It is then clear that:
\begin{gather}
\int^{\infty}_{x}t^r\Gamma\left({\alpha}/{\beta},t^\beta \right)dt
\notag\\
=
(r+1)^{-1}
\left. y^{\frac{r+1}{\beta}}\Gamma\left({\alpha}/{\beta},y \right)\right|^\infty_x
-(r+1)^{-1}
\int^{\infty}_{x}y^{\frac{r+1}{\beta}}\cdot
\left( -y^{\frac{\alpha}{\beta}-1}\text{e}^{-y}\right) dy,
\end{gather}

from which the result follows.


\bibliographystyle{agsm}
\bibliography{btnormalwork}

@PREAMBLE{ "\newcommand{\noopsort}[1]{} "# "\newcommand{\singleletter}[1]{#1} " }

@Article{allard07,
  Title                    = {A {N}ew {S}patial {S}kew-{N}ormal {R}andom {F}ield {M}odel},
  Author                   = {Allard, D. and Naveau, P.},
  Journal                  = {Communications in {S}tatistics-{T}heory and {M}ethods},
  Year                     = {2007},
  Number                   = {9},
  Pages                    = {1821-1834},
  Volume                   = {36},

  Publisher                = {Taylor \& Francis}
}

@Article{linearmixed05,
  Title                    = {Skew-normal linear mixed models},
  Author                   = {Arellano-Valle, R.B. and Bolfarine, H. and Lachos, V.H.},
  Journal                  = {Journal of {D}ata {S}cience},
  Year                     = {2005},
  Number                   = {4},
  Pages                    = {415-438},
  Volume                   = {3}
}

@Article{ayebo03,
  Title                    = {An asymmetric generalization of {G}aussian and Laplace laws},
  Author                   = {Ayebo, A. and Kozubowski, T. J.},
  Journal                  = {Journal of Probability and Statistical Science},
  Year                     = {2003},
  Number                   = {2},
  Pages                    = {187--210},
  Volume                   = {1}
}

@Article{azzalini85,
  Title                    = {A class of distributions which includes the normal ones},
  Author                   = {Azzalini, A.},
  Journal                  = {Scandinavian journal of statistics},
  Year                     = {1985},
  Pages                    = {171-178},
  Publisher                = {JSTOR}
}

@Article{azzalini08,
  Title                    = {Robust {L}ikelihood {M}ethods {B}ased on the {S}kew-t and {R}elated {D}istributions},
  Author                   = {Azzalini, A. and Genton, M.G.},
  Journal                  = {International {S}tatistical {R}eview},
  Year                     = {2008},
  Number                   = {1},
  Pages                    = {106-129},
  Volume                   = {76},

  Publisher                = {Wiley Online Library}
}

@Article{baker19,
  Title                    = {New survival distributions that quantify the gain from eliminating flawed components},
  Author                   = {Baker, R.},
  Journal                  = {Reliability {E}ngineering \& {S}ystem {S}afety},
  Year                     = {2019},
  Pages                    = {493--501},
  Volume                   = {185},

  Publisher                = {Elsevier}
}

@Article{barndorff78,
  Title                    = {Hyperbolic {D}istributions and {D}istributions on {H}yperbolae},
  Author                   = {Barndorff-Nielsen, O.},
  Journal                  = {Scandinavian {J}ournal of {S}tatistics},
  Year                     = {1978},
  Pages                    = {151--157},

  Publisher                = {JSTOR}
}

@Article{barndorff82,
  Title                    = {Normal {V}ariance-{M}ean {M}ixtures and z {D}istributions},
  Author                   = {Barndorff-Nielsen, O. and Kent, J. and S{\o}rensen, M.},
  Journal                  = {{I}nternational {S}tatistical {R}eview/{R}evue {I}nternationale de {S}tatistique},
  Year                     = {1982},
  Number                   = {2},
  Pages                    = {145-159},
  Volume                   = {50},

  Publisher                = {JSTOR}
}

@Article{bottazzi11,
  Title                    = {A new class of asymmetric exponential power densities with applications to economics and finance},
  Author                   = {Bottazzi, G. and Secchi, A.},
  Journal                  = {Industrial and Corporate Change},
  Year                     = {2011},
  Number                   = {4},
  Pages                    = {991--1030},
  Volume                   = {20},

  Publisher                = {Oxford University Press}
}

@Article{box62,
  Title                    = {A further look at robustness via {B}ayes's theorem},
  Author                   = {Box, G. E.P. and Tiao, G.C.},
  Journal                  = {Biometrika},
  Year                     = {1962},
  Number                   = {3},
  Pages                    = {419--432},
  Volume                   = {49},

  Publisher                = {JSTOR}
}

@Book{box73,
  Title                    = {Bayesian {I}nference in {S}tatistical {A}nalysis},
  Author                   = {Box, G. E.P. and Tiao, G. C.},
  Publisher                = {Addison {W}esley},
  Year                     = {1973},
  Volume                   = {1}
}

@Article{box64,
  Title                    = {An analysis of transformations},
  Author                   = {Box, G.E. P. and Cox, D.R.},
  Journal                  = {Journal of the Royal Statistical Society},
  Year                     = {1964},
  Number                   = {211},
  Pages                    = {1-43},
  Volume                   = {26}
}

@Book{deVries1894,
  Title                    = {Ueber halbe {G}alton-{C}urven als {Z}eichen discontinuirlicher {V}ariation},
  Author                   = {De Vries, H.},
  Publisher                = {Gebr{\"u}der {B}orntraeger},
  Year                     = {1894}
}

@Book{rent99,
  Title                    = {Regression: {M}odels, {M}ethods and {A}pplications},
  Author                   = {Fahrmeir, L. and Kneib, T. and Lang, S. and Marx, B.},
  Publisher                = {Springer {S}cience {\&} {B}usiness {M}edia},
  Year                     = {2013}
}

@Article{fernandez95,
  Title                    = {Modeling and {I}nference with $\upsilon$-{S}pherical {D}istributions},
  Author                   = {Fernandez, C. and Osiewalski, J. and Steel, M. F.J.},
  Journal                  = {Journal of the {A}merican {S}tatistical {A}ssociation},
  Year                     = {1995},
  Number                   = {432},
  Pages                    = {1331--1340},
  Volume                   = {90},

  Publisher                = {Taylor \& Francis Group}
}

@Article{ferreira06,
  Title                    = {A constructive {R}epresentation of {U}nivariate {S}kewed {D}istributions},
  Author                   = {Ferreira, J.T.A.S. and Steel, M.F.J.},
  Journal                  = {Journal of the {A}merican {S}tatistical {A}ssociation},
  Year                     = {2006},
  Number                   = {474},
  Pages                    = {823-829},
  Volume                   = {101},

  Publisher                = {Taylor \& Francis}
}

@Book{Gr<3dstheyn,
  Title                    = {Table of Integrals, Series and Products},
  Author                   = {Gradshteyn,I.S. and Ryhzhik,I.M.},
  Editor                   = {Jeffrey,A. and Zwillinger,D.},
  Publisher                = {Academic Press},
  Year                     = {2007}
}

@Article{hansen94,
  Title                    = {Autoregressive {C}onditional {D}ensity {E}stimation},
  Author                   = {Hansen, B. E.},
  Journal                  = {International {E}conomic {R}eview},
  Year                     = {1994},
  Number                   = {3},
  Pages                    = {705-730},
  Volume                   = {35},

  Publisher                = {JSTOR}
}

@Article{deHel1909,
  Title                    = {Sulla rappresentazione analitica delle curve statistiche},
  Author                   = {de Helguero, F.},
  Journal                  = {Giornale degli {E}conomisti},
  Year                     = {1909},
  Number                   = {20},
  Pages                    = {241-265},
  Volume                   = {38},

  Publisher                = {JSTOR}
}

@Article{jones15,
  Title                    = {On {F}amilies of {D}istributions with {S}hape {P}arameters},
  Author                   = {Jones, M.C.},
  Journal                  = {International {S}tatistical {R}eview},
  Year                     = {2015},
  Number                   = {2},
  Pages                    = {175-192},
  Volume                   = {83},

  Publisher                = {Wiley Online Library}
}

@Article{jones04,
  Title                    = {Families of {D}istributions {A}rising from {D}istributions of {O}rder {S}tatistics},
  Author                   = {Jones, M.C.},
  Journal                  = {Test},
  Year                     = {2004},
  Number                   = {1},
  Pages                    = {1-43},
  Volume                   = {13},

  Doi                      = {https://doi.org/10.1007/BF02602999},
  Publisher                = {Springer}
}

@Article{jones11,
  Title                    = {On parameter orthogonality in symmetric and skew models},
  Author                   = {Jones, M.C. and Anaya-Izquierdo, K.},
  Journal                  = {Journal of {S}tatistical {P}lanning and {I}nference},
  Year                     = {2011},
  Number                   = {2},
  Pages                    = {758-770},
  Volume                   = {141},

  Publisher                = {Elsevier}
}

@Article{bayf95,
  Title                    = {Bayes {F}actors},
  Author                   = {Kass, R. E. and Raftery, A. E.},
  Journal                  = {Journal of the {A}merican {S}tatistical {A}ssociation},
  Year                     = {1995},
  Number                   = {430},
  Pages                    = {773-795},
  Volume                   = {90},

  Publisher                = {Taylor \& Francis}
}

@Article{komunjer07,
  Title                    = {Asymmetric power distribution: {T}heory and applications to risk measurement},
  Author                   = {Komunjer, I.},
  Journal                  = {Journal of applied econometrics},
  Year                     = {2007},
  Number                   = {5},
  Pages                    = {891--921},
  Volume                   = {22},

  Publisher                = {Wiley Online Library}
}

@Article{ley14flex,
  Title                    = {Flexible modelling in statistics: past, present and future},
  Author                   = {Ley, C.},
  Journal                  = {arXiv preprint arXiv:1409.6219},
  Year                     = {2014}
}

@Article{baylinmod18,
  Title                    = {A flexible class of parametric distributions for {B}ayesian linear mixed models},
  Author                   = {Maleki, M. and Wraith, D. and AArellano-Valle, R. B.},
  Journal                  = {{TEST}},
  Year                     = {2018},
  Number                   = {2},
  Pages                    = {1-22},
  Volume                   = {28},

  Publisher                = {Springer}
}

@Article{mclachlan2000,
  Title                    = {Finite {M}ixture {M}odels},
  Author                   = {McLachlan, G. J. and Lee, S. X. and Rathnayake, S. I.},
  Journal                  = {Annual {R}eview of {S}tatistics and {i}ts {A}pplication},
  Year                     = {2000},
  Pages                    = {355-378},
  Volume                   = {6},

  Publisher                = {Annual Reviews 4139 El Camino Way, PO Box 10139, Palo Alto, California 94303~…}
}

@Article{nadarajah12,
  Title                    = {On the {C}haracteristic {F}unction for {A}symmetric {E}xponential {P}ower {D}istributions},
  Author                   = {Nadarajah, S. and Teimouri, M.},
  Journal                  = {Econometric {R}eviews},
  Year                     = {2012},
  Number                   = {4},
  Pages                    = {475--481},
  Volume                   = {31},

  Publisher                = {Taylor \& Francis}
}

@Article{naveau05,
  Title                    = {A skewed {K}alman filter},
  Author                   = {Naveau, P. and GeGenton, M.. and Shen, X.},
  Journal                  = {Journal of {m}ultivariate {A}nalysis},
  Year                     = {2005},
  Number                   = {2},
  Pages                    = {382-400},
  Volume                   = {94},

  Publisher                = {Elsevier}
}

@Book{nelsen07,
  Title                    = {An {I}ntroduction to {C}opulas},
  Author                   = {Nelsen, R. B.},
  Publisher                = {Springer {S}cience \& {B}usiness {M}edia},
  Year                     = {2007}
}

@Article{nonlinearmixed19,
  Title                    = {Nonlinear mixed-effects models with scale mixture of skew-normal distributions},
  Author                   = {Pereira, M. A. A. and Russo, C. M.},
  Journal                  = {Journal of {A}pplied {S}tatistics},
  Year                     = {2019},
  Number                   = {9},
  Pages                    = {1602-1620},
  Volume                   = {46},

  Publisher                = {Taylor \& Francis}
}

@Article{rubio14,
  Title                    = {Inference in {T}wo-piece {L}ocation-{S}cale {M}odels with {J}effreys priors},
  Author                   = {Rubio, F. J. and Steel, M.F.J.},
  Journal                  = {Bayesian {A}nalysis},
  Year                     = {2014},
  Number                   = {1},
  Pages                    = {1--22},
  Volume                   = {9},

  Publisher                = {International Society for Bayesian Analysis}
}

@Article{genhypscott11,
  Title                    = {Moments of the generalized hyperbolic distribution},
  Author                   = {Scott, D. J. and W{\"u}rtz, D. and Dong, C. and Tran, T. T.},
  Journal                  = {Computational {S}tatistics},
  Year                     = {2011},
  Number                   = {3},
  Pages                    = {459-476},
  Volume                   = {26},

  Publisher                = {Springer}
}

@Article{stacy62,
  Title                    = {A {G}eneralization of the {G}amma {D}istribution},
  Author                   = {Stacy, E. W.},
  Journal                  = {The Annals of {M}athematical {S}tatistics},
  Year                     = {1962},
  Number                   = {3},
  Pages                    = {1187--1192},
  Volume                   = {33},

  Publisher                = {Institute of Mathematical Statistics}
}

@Article{subbotin23,
  Title                    = {On the {L}aw of {F}requency of {E}rror},
  Author                   = {Subbotin, M. T.},
  Journal                  = {Mathematicheskii {S}bornik},
  Year                     = {1923},
  Number                   = {2},
  Pages                    = {296--301},
  Volume                   = {31},
  Publisher                = {Российская академия наук, Математический институт им. ВА Стеклова Российской~…}
}

@Article{zhu09,
  Title                    = {Properties and estimation of asymmetric exponential power distribution},
  Author                   = {Zhu, D. and Zinde-Walsh, V.},
  Journal                  = {Journal of Econometrics},
  Year                     = {2009},
  Number                   = {1},
  Pages                    = {86--99},
  Volume                   = {148},

  Publisher                = {Elsevier}
}

@InCollection{johnson94,
  author    = {Johnson, N. and Kotz, S. and Balakrishnan, N.},
  title     = {Continuous univariate distributions},
  publisher = {Wiley},
  year      = {1994},
  volume    = {2},
}

\end{document}